\documentclass[aps,twocolumn,showpacs]{revtex4}
\usepackage{epsfig}
\usepackage{graphicx}

\begin{document}

\title{Quantum Fingerprinting with a Single Particle}
\author{S. Massar}

\altaffiliation[Also at: ]{Ecole Polytechnique, C.P. 165, 
Universit\'e Libre de Bruxelles, 1050
Brussels, Belgium}
\affiliation{Service de Physique Th\'eorique, Universit\'e 
Libre de Bruxelles, C.P. 225,
Bvd. du Triomphe, 1050 Bruxelles, Belgium}
\date{\today}
\begin{abstract}
We show that the two slit experiment in which a single quantum
particle interferes with itself can be interpreted as a quantum
fingerprinting protocol: the interference pattern exhibited by the
particle contains information about the environment it encountered
in the slits  which would require much more communication to learn
classically than is required quantum mechanically.
An extension to the case
where the particle has many internal degrees of freedom is suggested and its
interpretation is discussed in detail. A possible experimental
realization is proposed.

 \end{abstract}
\pacs{ 03.65.Ta, 03.67.Hk, 42.50.Xa}
\maketitle

\section{Introduction}

As first understood by de Broglie\cite{deB}, a single quantum particle should
exhibit both particle and wave like properties. The simplest way to
illustrate the wave-particle duality of quantum mechanics is the two
slit experiment in which a single particle can follow one of  two
paths and then interfere with itself. 
When a single particle is sent through the
setup a single click is registered, as is expected of individual
particles. But upon accumulating statistics an interference pattern
emerges, putting in evidence their wave like behavior. 
The interference pattern only occurs if it is impossible to
know which path the particle took. The two slit experiment is one of the
fundamental Gedanken-Experimenten discussed by the fathers of quantum
mechanics\cite{Heis}. It has by now been realized with a wide variety of physical
systems, such as  electrons\cite{el}, neutrons\cite{neutron}, 
photons\cite{photon}, single atoms\cite{atom} and even
simple molecules\cite{molecule}. 

The aim of this paper is to reexamine the two slit experiment in the
light of recent advances in quantum information theory, and in
particular the quantum fingerprinting protocol introduced in
\cite{BCWdW}. We will show that to reproduce the interference pattern
produced by a single particle would require much more classical
communication than is necessary quantum mechanically. This allows one
to put constraints on possible classical theories which could underly
quantum mechanics.

The situation we will consider is the following (see figure 1). 
A single particle
impinges on a beam splitter. If the particle is transmitted it goes to
one party, Alice. If the particle is reflected it goes to another
party, Bob. Thus the state is  
$$ \left( |1_A\rangle + |1_B\rangle \right ) /\sqrt{2} $$ 
where the subscript $A$ or $B$ denotes where the particle is.
Alice and Bob also receive a classical input. Alice receives as input
$x$ and Bob receives as input $y$. Alice and Bob then carry out a
local unitary transformation on the particle which depends on their
input. Thus the state becomes  
$$  \left( U_A(x) |1_A\rangle + U_B(y)|1_B\rangle  \right ) / \sqrt{2}\ .$$
We consider the possibilities that the particle either has or does
not have internal degrees of freedom. If the particle does not have
internal degrees of freedom, then the transformations $U_A(x)= e^{i
\phi_A(x)}$ and $U_B(x)= e^{i \phi_B(x)}$ 
are simply phases. If the particle does have internal degrees of
freedom, then the transformations $U_A$ and $U_B$ can act on the
internal degrees of freedom. In general they transform the state to 
$$
U_A(x)|1_A\rangle = \sum_{i=1}^d \alpha_i(x) |i_A\rangle 
\quad , \quad
 U_B(y)|1_B\rangle = \sum_{i=1}^d \beta_i(y) |i_B\rangle
$$
where $d$ is the dimensionality of the internal Hilbert space.
These internal degrees of freedom could for instance be the spin,
momentum, or energy of the particle. Then, exactly as in the
traditional two slit experiment,  the particle goes to a third party,
which we call the Referee, who makes the trajectories coming from
Alice interfere with the trajectories coming from Bob. 
We will suppose that this is done using a beam splitter which combines
the two incoming beams as: 
$$
|i_A\rangle \to ( |i_E \rangle + |i_N\rangle ) / \sqrt{2}
\quad , \quad
|i_B\rangle \to ( |i_E \rangle - |i_N\rangle ) / \sqrt{2} \ ,
$$
to yield the state
$$ {1 \over 2} \sum_{i=1}^d
(\alpha_i(x) + \beta_i(y))  |i_E \rangle + ( \alpha_i(x) - \beta_i(y)
)  |i_N\rangle  
\ .$$
The  Referee then measures the particle in the basis $ |i_E \rangle
,|i_N\rangle $. We have denoted the output ports of the final beam
splitter by $E$ (for ``Equal'') and $N$ (for ``Not Equal'')
because if $U_A = U_B$ the particle exits by port $E$, whereas if $U_A=-U_B$
the particle exits by port $N$. We shall show that if $U_A$ and $U_B$
are adequately chosen the
interference pattern of the particle with itself allows the Referee to
learn information about the inputs $x$ and $y$ which he could only
learn
 using much more classical communication than the amount of
quantum communication that takes
place in the two slit experiment.

In summary, when the particle does not have internal degrees of
freedom this is the well known two slit experiment, 
although presented in a slightly different way. When the
particle does have internal degrees of freedom then this is a
straightforward generalization of the two slit experiment.  

\section{Classical and Quantum Fingerprinting}

Before proceeding further we first
recall what is known about classical and quantum fingerprinting
protocols. In the protocols we consider there are three
parties, Alice, Bob and a Referee. Alice gets as input an $n$ bit
string $x \in \{ 0,1\}^n$. Bob gets as input an $n$ bit string $y\in
\{0,1\}^n$. The aim is for the Referee to decide whether $x=y$ or $x\neq
y$. The Referee can make mistakes, but the probability of a mistake
must be less than $\epsilon$ for some $1/2 > \epsilon >0$. (By repeating a protocol a constant times, the error probability can be made
exponentially small if $\epsilon$ is not very close to $1/2$ to
start. For instance $\epsilon = 1/3$ is sufficient.). The aim is for
the Referee to reach a decision while minimizing the amount of
communication between the parties.  There are several different
protocols which depend on the resources used by the parties. Let us
enumerate these different possibilities. 

\begin{enumerate}
\item
Alice, Bob and the Referee do not have access to any randomness and
must use a deterministic protocol. Then $O(n)$ bits of classical
communication
are necessary\cite{KN}. 

\item  One way classical communication from Alice to the Referee and 
one way classical
communication from Bob to the Referee is allowed 
(the "simultaneous message passing"
model introduced by Yao\cite{Y}). Alice and Bob have access to a local
source of randomness. Then $O(\sqrt{n})$ bits of
communication are necessary\cite{A,NS,BK}. 

\item One way quantum communication from Alice to the Referee and 
one way quantum
communication from Bob to the Referee is allowed 
(the "quantum simultaneous message passing" model). Then
$O(\log{n})$ qubits of communication are sufficient\cite{BCWdW}. 

\item  Classical communication between Alice, Bob and the Referee is
allowed in all directions. Alice and Bob have access to a local source of
randomness. Then $O(\log n)$ bits of communication are
sufficient\cite{KN}. 

\item  One way classical communication from Alice to the Referee and 
one way classical
communication from Bob to the Referee is allowed 
(the "simultaneous message passing"
model). Alice and Bob share $O(\log n)$ random bits. Then a protocol
exists in which Alice sends
$O(1)$ bits to the Referee and Bob sends $O(1)$ bits to the Referee.  

\end{enumerate}

\section{Quantum Fingerprinting Using a Single Particle with No
Internal Degrees of Freedom}

Let us first show that even when the particle has no internal degrees
of freedom, the two slit experiment realizes a non trivial quantum
communication protocol.
The simplest case is when Alice's input consists of a
single bit $x=0,1$ and Bob's input consists of a single bit $y=0,1$.
Then the quantum strategy is simply:
$$
|1_A\rangle \to e^{i \pi x} |1_A\rangle \quad ; \quad
|1_B\rangle \to e^{i \pi y} |1_B\rangle \ .
$$
This strategy solves the quantum fingerprinting problem with no error
and with a single
ebit of communication since the Hilbert space of the particle is two
dimensional. Classically, in the simultaneous message passing model, 
it is of course necessary for both parties to
send their input to the referee, ie. two bits of classical
communication are required. Note that this is the case irrespective of
whether the parties have shared randomness.

We have generalized the above protocol to the case where 
the inputs $x,y \in \{0,1,2\}$ are trits. The quantum strategies are 
$$|1_A\rangle \to e^{i2 \pi x / 3} |1_A\rangle
\quad , \quad 
|1_B\rangle \to e^{i2 \pi y / 3} |1_B\rangle \ .$$
One easily checks that with such a strategy the Referee makes no error
when $x=y$ and will recognize that $x \neq y$ with probability
$3/4$. Thus, if one averages over the inputs, the average error
probability is $1/6$. 

In order to  compare this 
with classical fingerprinting in the simultaneous message
passing model, we suppose that Alice can send one trit of
classical information to the Referee, 
whereas Bob is limited and can only send one
bit to the Referee. 
We will show that this is not enough communication to reproduce
the quantum protocol. Let us first consider deterministic
strategies. If Alice sends one trit of information, she can tell the
Referee what is her input.  On the other hand with one bit of
information Bob can only divide his inputs into two sets. One easily
checks that there are necessarily at least two input pairs ($x,y$) (out of
9 possible pairs) for which the Referee will make a mistake. Thus
averaged over the inputs the error probability is at least $2/9$,
which is greater than in the quantum protocol. 

Note that in this case shared randomness does not
allow the parties to do better than the quantum protocol. Indeed
shared randomness can be viewed as allowing  the parties to randomly
choose a deterministic strategy. Since all
deterministic strategies have an error probability of at least $2/9$,
the same is true of any probability distribution over deterministic
strategies.  

It is remarkable how economical the quantum protocol is compared to
the classical protocol: it uses a single qubit of communication, and
does better than any classical protocol in the simultaneous message
passing model in which a bit and a trit are communicated. To do better
than the quantum protocol two trits must be communicated. 

\section{Quantum Fingerprinting Using a Single Particle with Internal
Degrees of Freedom}

We now generalize the above protocols to the case
where the particle has
internal degrees of freedom.
The inputs are taken to be $n$ bit strings. Let $E$ be an error
correcting code which encodes words of length $n$ into words of length
$m$, and such that if two words are distinct, then the Hamming
distance between the encoded words is at least $t$.  
There exist such codes with the property that for sufficiently large
$n$, $m\leq \mu n$ and $t\geq \nu m$ where $\mu >1$ and $\nu >0$
are independent of  $n$. For instance in the case of 
Justesen codes $\nu =  { 1 \over 10} - {1 \over 15
\mu}$ for any $\mu > 2$ \cite{Justesen}.
We denote by $e(x)$ the encoded version of input $x$. The $i$'th bit
of $e(x)$ is denoted $e_i(x)$. 
The encoding operations carried out by Alice and Bob are 
\begin{eqnarray}
U_A(x)|1_A\rangle &=& {1 \over \sqrt{m}} \sum_{i=1}^{m}
(-1)^{e_i(x)} |i_A\rangle\ ,\nonumber\\
U_B(y)|1_B\rangle &=& {1 \over \sqrt{m}} \sum_{i=1}^{m}
(-1)^{e_i(y)} |i_B\rangle\ .\nonumber
\end{eqnarray}
When the particle is received by the referee, he lets the paths coming
from Alice and the paths coming from Bob interfere as described above
and measures whether the particle is in the subspace spanned by the
$|i_E\rangle$ states or in the subspace spanned by the $|i_N\rangle$
states. If $x=y$ then the particle necessarily is in one of the
$|i_E\rangle$ states. On the other hand if the inputs are different,
then the probability that the particle is in one of the $|i_N\rangle$
states is at least $t/m \geq \nu >0$. Thus the probability of error is
less than $1 - \nu$. By repeating the protocol $k$ times, with $k \geq
\log \epsilon / \log
(1 - \nu)$, one can make the error probability less than $\epsilon$. 

\section{Interpretation}

We now turn to the interpretation of the above interference experiments. 
To this end
we must compute the quantum and classical capacities of the
communication channels used, and we must compute how much shared
randomness exists between the parties. 
One of the aims of this discussion is to show that
these experiment can constrain the properties of any classical theory
which could underly quantum mechanics, but in the context of a single
particle, in a similar way that EPR type experiments can constrain the
properties of any classical theory which could underly quantum
mechanics in the context of two entangled particles. 

First we characterize the quantum experiment. To this end
we need to compute how many qubits are transmitted. When Alice and Bob
send the particle to the Referee, the particle belongs to a Hilbert
space of dimension $2m$ (since there are $m$ paths from Alice to the
Referee and $m$ paths from Bob to the Referee). The total capacity of
the channels used to communicate to the referee is therefore $1 + \log
m$ qubits.

Let us now compare this quantum protocol with classical protocols
which achieve the same fingerprinting. First of all we note that since
the communication from Alice and Bob to the Referee uses $1+ \log m$
qubits,  Holevo's theorem\cite{Holevo} implies that at most $1 + \log m$
bits of classical communication could be transmitted during this
phase.  
In order to carry out fingerprinting on $n$ bit inputs, the results
stated above imply that classically either communication in all
directions should be allowed (ie. one is not restricted to
simultaneous message passing ), or the parties have prior shared
randomness.  

Concerning the first point, the only communication that takes place
between the parties after the inputs is received is from Alice to the
Referee and from Bob to the Referee. Hence it is natural to argue that
no other communication between the parties is possible in the
classical case either. It is possible to strengthen this argument by
appealing to special relativity. Indeed let us suppose that the Referee is
mid way between Alice and Bob in such a way that the distance between
Alice and Bob is $L$, and the distance between Alice or Bob and the
Referee is $L/2$. We can suppose that Alice and Bob receive their
inputs $x$ and $y$ at time $t=0$, and that the referee must provide an
output before time $t= 3L/2c$. This implies that if no information can
travel faster than the speed of light, then one must necessarily be in
the simultaneous message passing model. 

Let us now consider whether the parties have shared randomness. This
shared randomness could presumably only be produced when the particle
is sent to Alice and Bob. During this phase the state of the particle
can be written as  
\begin{equation}
(|1_A\rangle |vac_B\rangle+ |vac_A\rangle|1_B\rangle)/\sqrt{2}
\label{ent}
\end{equation}
 where $|vac\rangle$ is the vacuum state. Hence this is an entangled
state with one unit of entanglement (one ebit). By measuring whether
or not they have the particle, Alice and Bob can generate one shared
random bit. However a single shared random bit cannot
decrease the amount of communication required for fingerprinting in
the simultaneous message passing model by more than a factor 
2\cite{explain}.

In summary quantum fingerprinting with a single particle seems to be
incompatible with a classical description of the particle. If one
insists on keeping a classical description then the underlying
classical theory must have one of the following (surprising) properties
\begin{enumerate}
\item
 A quantum particle following $m$ different paths carries $O(\sqrt{m})$
bits of classical information (which is exponentially more than the
amount of classical information obtained by measuring in which path is
the particle). 
\item A single particle in the state eq. (\ref{ent}) carries with it
$O(\log m)$ bits of shared randomness along the two paths taken by the
particle (whereas quantum mechanics predicts that only one bit of
shared randomness could be obtained by measuring the state (\ref{ent}). 
\item  The classical theory allows $O(\log m)$ bits of
superluminal communication if the particle follows $m$ different
paths.
\end{enumerate}

The best known classical theory which reproduces the behavior of a
single quantum particle is Bohm's theory\cite{Bohm} in which the particle
follows a well defined classical trajectory, but is accompanied by a
"pilot wave" which determines the trajectory the particle must
take. Bohm's theory realizes the first possibility outlined above
since the pilot wave is a classical description of the quantum wave
function and therefore contains $O(m)$ bits of classical
information. The possibility of realizing quantum fingerprinting with
a single particle shows that it is impossible to imagine a
"compressed" version of Bohm's theory in which the pilot wave carries
less than $O(\sqrt{m})$ bits. 

Since the above discussion makes appeal to causality, it is
interesting to compare it to Bell's appeal to causality when
interpreting EPR type experiments\cite{Bell}. 
Bell showed that "local realistic
theories" are incompatible with quantum mechanics. These are 
classical theories which respect special
relativity and which allow
the entangled particles to carry unlimited shared randomness with
them.
Many experiments on Bell type correlations have been carried
out over the years, but no completely conclusive experiment has been
realized.  
Quantum fingerprinting with a single particle is weaker than Bell's
tests of quantum correlations since it
does not allow one to disprove local realistic
theories. However it does allow bounds to be put on the minimum amount of 
randomness which must be carried by a single 
quantum particle, or on the minimum
amount of superluminal communication which must take place between the
parties.

\section{ Experimental realization of
Quantum Fingerprinting with a single particle}

We will now show that realising a two slit experiment with a
single particle with a large number (of order $10^3$)  
internal degrees of freedom
seems possible with present technology, but that more theoretical work
is necessary in order to establish whether non trivial fingerprinting
takes place under these conditions. The setup is illustrated in figure
1.
The particle is taken to be a photon  produced by a 
mode-locked laser. Such lasers
produce trains of light pulses with very long coherence
lengths (up to $10^{16}$ pulses may be coherent\cite{Th}). Note that
it is essential that all the pulses
in the train are coherent otherwise one is not dealing with a pure
state.
The light produced by the mode locked laser is attenuated so that
on average less than  a
single photon is present in the pulse train.
If a single photon is present, then the state of light is
$$
\psi = {d^{-1/2}} \sum_{i=1}^{d} |i\rangle \ .$$
This train of pulses then impinges on a beam splitter. The transmitted
beam is sent to Alice, the reflected beam to Bob. Thus the state
becomes
$$
\psi = {d^{-1/2} \over \sqrt{2}}\sum_{i=1}^{d} |i_A\rangle + 
|i_B\rangle \ .$$
Alice and Bob can encode their input into the train of pulses by
using a phase modulator. Note that the large value for $d$ mentioned above is
certainly not realistic, and imperfections, such as dark counts from
detectors, will limit $d$ to much smaller values.

By coupling the
pulses into optical fibers, it should be possible to separate Alice
and Bob by distances of several kilometers. If the
pulses are separated by 1ns, then with Alice and Bob separated by 10
km, corresponding to a time separation of $3\mu s$, 
it should be possible to have $d$ of order
$10^3$ while keeping Alice and Bob's encoding operations spatially separated.
This value for $d$ seems much more realistic, but is still impressive
if one notes that it corresponds to approximately $10$ qubits.

Let us compare these figures with the theoretical predictions.
In \cite{BK} it is proven that in the simultaneous message
passing model, if Alice and Bob have no
shared randomness, if the referee uses a deterministic decision
strategy, and if the error probability is less than $1 \%$, then the number
$a$ 
of bits of communication sent by Alice to the referee, and the number
$b$ of bits of communication sent by Bob to the referee, must obey $a
b \geq n /400$. This implies that one at least of $a$ and $b$ must be
greater than $\sqrt{n}/20$. With a single bit of shared randomness,
Alice and Bob must send at least half this amount of
communication\cite{explain}, ie. at least $\sqrt{n}/40$ bits.

Let us suppose that in the protocol described above Alice and Bob use
Justesen codes with the parameter $\mu=2$, which implies $\nu = 1/15$.
Thus the number of internal degrees of freedom of the particle must be
$2n$, and the number of qubits communicated by each party in a
single run of the protocol is $1 + \log_2 n$.
In order for the error probability to be less than $0.01$, the parties
must
repeat the protocol $k =67 \geq \ln (0.01) / \ln (1- \nu)$ times.
The total amount of quantum communication sent by each party is $k (1
+ \log_2 n)$. This must be smaller than $\sqrt{n} /40$ which is
satisfied if
 $n$ is larger than about $ 10^{10}$. 

As mentioned above it is  
probably impossible to reach such large values of $n$ in a
realistic experiment and the  
The value of
$d=10^3$ suggested above seems much more realistic. However in this
case it is unclear whether one is doing non trivial fingerprinting.
Nevertheless it may be possible, by optimising the proofs, 
to decrease the
required values of $d$ significantly. 
In fact it is not unreasonable
to conjecture that non trivial quantum fingerprinting is possible for all
values of $d$, since we know that it is possible when $n=1$ 
(particle with no internal degrees of freedom), and it is possible for
sufficiently large $d >  10^{10}$.

\section{Conclusion}

In summary we have shown that recent theoretical
advances in quantum communication
allow a new interpretation of one of the most famous experiments of
quantum mechanics, namely the interference of a single particle with
itself. Indeed the two slit experiment can be
interpreted as a quantum fingerprinting experiment which requires
more classical communication than is used quantum mechanically.
A generalisation of quantum fingerprinting with a single particle 
to the case
where the particle has internal degrees of freedom is proposed. 
The
possibility of realising such an experiment using
a photon in a train of many coherent pulses is also discussed. 
The
interpretation of these experiments is discussed in detail.

{\bf Acknowledgements:} I would like to thank Harry Burhman, Hein Ro\"erig
and Louis-Philippe Lamoureux for helpfull discussions.  Financial support 
from the Communaut\'e
Fran\c{c}aise de Belgique under grant ARC 00/05-251, from the IUAP
programme of the Belgian government under grant V-18 and from  
the EU through project RESQ (IST-2001-37559) is gratefully acknowledged.

\begin{figure}[htb]
{
\unitlength=1.000000pt
\begin{picture}(190,120)(0.00,0.00)
\put(95.00,63.75){\line(-1,0){9.50}}
\put(95.00,58.25){\line(0,1){5.50}}
\put(85.25,58.25){\line(1,0){9.75}}
\put(85.25,63.75){\line(0,-1){5.50}}
\put(155.00,64.00){\line(-1,0){9.75}}
\put(155.00,58.75){\line(0,1){5.25}}
\put(145.00,58.75){\line(1,0){10.00}}
\put(145.00,64.25){\line(0,-1){5.50}}
\put(115.00,14.75){\framebox(10.00,12.25){$\phi_B$}}
\put(115.00,95.00){\framebox(9.75,11.00){$\phi_A$}}
\qbezier(175.25,36.75)(179.50,37.25)(179.00,41.00)
\qbezier(175.25,45.75)(179.50,45.75)(179.00,40.50)
\qbezier(175.25,77.00)(179.50,77.00)(179.00,81.00)
\qbezier(175.25,86.00)(179.25,86.00)(179.25,80.75)
\put(175.00,46.00){\line(0,-1){9.50}}
\put(175.00,86.00){\line(0,-1){9.25}}
\qbezier(164.75,51.00)(164.75,41.00)(174.75,41.00)
\qbezier(154.75,61.00)(164.75,61.00)(164.75,51.00)
\qbezier(164.75,71.00)(164.75,81.00)(174.75,81.00)
\qbezier(154.75,61.00)(164.75,61.00)(164.75,71.00)
\qbezier(134.75,41.00)(134.75,61.00)(144.75,61.00)
\qbezier(124.75,21.00)(134.75,21.00)(134.75,41.00)
\qbezier(134.75,81.00)(134.75,61.00)(144.75,61.00)
\qbezier(124.75,101.00)(134.75,101.00)(134.75,81.00)
\qbezier(104.75,41.00)(104.75,21.00)(114.75,21.00)
\qbezier(94.75,61.00)(104.75,61.00)(104.75,41.00)
\qbezier(104.75,81.00)(104.75,101.00)(114.75,101.00)
\qbezier(94.75,61.00)(104.75,61.00)(104.75,81.00)
\put(84.75,61.00){\rule{0.00\unitlength}{0.00\unitlength}}
\qbezier(84.75,61.00)(74.75,61.00)(74.75,71.00)
\put(64.75,61.00){\line(1,0){20.00}}
\put(45.00,61.00){\line(1,0){8.75}}
\put(190.00,41.00){\makebox(0.00,0.00)[l]{{\em Equal}}}
\put(190.00,81.25){\makebox(0.00,0.00)[l]{{\em Not Equal}}}
\put(120.00,5.00){\makebox(0.00,0.00){BOB}}
\put(120.00,115.00){\makebox(0.00,0.00){ALICE}}
\put(190.00,60.75){\makebox(0.00,0.00){REFEREE}}
\put(184.75,41.00){\makebox(0.00,0.00){D}}
\put(184.75,81.25){\makebox(0.00,0.00){D}}
\put(53.75,55.25){\framebox(11.00,10.25){A}}
\put(147.00,70.25){\makebox(0.00,0.00)[l]{C}}
\put(87.00,70.25){\makebox(0.00,0.00)[l]{C}}
\put(0.00,56.00){\framebox(45.00,11.00){M L Laser}}
\end{picture}}
\caption{Quantum Fingerprinting using a single photon with many internal
degrees of freedom. The mode locked laser (M L Laser)
produces a train of
pulses. The train is coupled into an optical fiber. An 
attenuator (A) decreases the intensity so
that the train contains less than a
single photon. A coupler (C)
sends with amplitude $1/\sqrt{2}$ the pulses to Alice, and with
amplitude $1/\sqrt{2}$ the pulses to Bob. Alice and Bob put phases on
the successive pulses using their phase modulators $\phi_A$ and
$\phi_B$. The pulses are then sent back to the referee who recombines
them using a coupler. If the phases put by Alice and Bob are equal,
the photon (if one is present) 
exits by the {\em Equal} port. If they are opposite, they
exit by the {\em Not Equal} port. Finally the photon is detected by
single photon detectors (D).
}\label{Lfig2}
\end{figure}
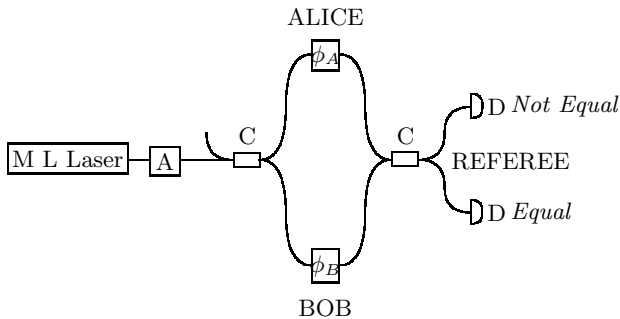


\begin{thebibliography}{99}
\bibitem{explain} Suppose that
Alice and Bob do not have the shared random bit. Then Alice could 
send
the information she would send if the bit had value 0 as well as the
information she would send if the bit had value 1. Similarly for
Bob. This protocol is at least as good as the protocol with the
shared random bit, but uses twice as much conversation.
\end{thebibliography}
\end{document}